\documentclass[conference]{IEEEtran}
\usepackage{cite}
\usepackage{graphicx}
\usepackage{tabto}
\usepackage{float}

\usepackage{lineno,hyperref}

\ifCLASSINFOpdf
\else
\fi
\begin{document}
%
\title{HTCC: Haskell to Handel-C Hardware Compiler}

\title{{\large \textcopyright 2016 IEEE. Personal use of this material is permitted. Permission from IEEE must be obtained for all other uses, in any current or future media, including reprinting/republishing this material for advertising or promotional purposes, creating new collective works, for resale or redistribution to servers or lists, or reuse of any copyrighted component of this work in other works. DOI: 10.1109/DSD.2016.24\\ .\\ A. Ablak and I. Damaj, HTCC: Haskell to Handel-C Compiler, The $19^{th}$ EUROMICRO Conference on Digital System Design, IEEE, Limassol, Cyprus, August 31\textendash September, 2016. P 192\textendash199. \\  \url{https://doi.org/10.1109/DSD.2016.24}}\\ HTCC: Haskell to Handel-C Hardware Compiler}

\author{\IEEEauthorblockN{Ahmed B. Ablak and Issam Damaj}
\IEEEauthorblockA{Electrical and Computer Engineering Department\\
American University of Kuwait\\
Salmiya, Kuwait\\
Email: \{s00015070, idamaj\}@auk.edu.kw}}


%


\maketitle

\begin{abstract}
Functional programming languages, such as Haskell, enable simple, concise, and correct-by-construction hardware development. HTCC compiles a subset of Haskell to Handel-C language with hardware output. Moreover, HTCC generates VHDL, Verilog, EDIF, and SystemC programs. The design of HTCC compiler includes lexical, syntax and semantic analyzers. HTCC automates a transformational derivation methodology to rapidly produce hardware that maps onto Field Programmable Gate Arrays (FPGAs) . HTCC is generated using ANTLR compiler-compiler tool and supports an effective integrated development environment. This paper presents the design rationale and the implementation of HTCC. Several sample generations of first-class and higher-order functions are presented. In-addition, a compilation case-study is presented for the XTEA cipher. The investigation comprises a thorough evaluation and performance analysis. The targeted FPGAs include Cyclone II, Stratix IV, and Virtex-6 from Altera and Xilinx.
\end{abstract}


%
\IEEEpeerreviewmaketitle

\section{Introduction}
FPGAs are famous and widely used reconfigurable computing (RC) systems. FPGAs have become very popular in research and industrial applications in different fields, such as, security, signal processing etc. FPGAs evolved from being limited in functionality and speed to become high-performance processors. Example FPGAs include Stratix from Altera and Virtex from Xilinx \cite{c0,c00}. The flexibility of FPGAs, that are sometimes described as seas-of-gates, enable the development of software paradigms to rapidly reconfigure hardware almost instantly.  

Recently, there has been considerable focus on the development of high-level synthesis (HLS) and rapid prototyping hardware/software co-design tools. The targets of co-design tools are high design productivity, simplicity, reduced time-to-prototype, correctness, to name a few. Co-design tools include converting algorithmic behaviors into digital circuits that can map onto FPGAs. High-level co-design tools are currently beyond behavioral VHDL and the other standard tools. The area witnessed the emergence of programming languages and tools such as Handel-C \cite{c8}, SystemC \cite{c6}, Matlab HDL Coder, LabVIEW, etc. All the modern co-design tools enable the integration and partitioning of computations into communicating hardware and software subsystems.

Handel-C is a high-level language with hardware output. Handel-C is based on ANSI C; it is extended to the theory of communication sequential processes (CSP) and the concurrent programming language (OCCAM) \cite{c7}. Moreover, Handel-C has the ability to provide both parallel and sequential implementations. Handel-C can target different FPGA types. Recent research effort has been on automating hardware generation to target Handel-C and hardware in general starting from functional specifications, such as, Haskell \cite{c9,c1,c10,Damaj2007a}. 

Haskell is a purely functional programming language that utilizes functions to construct programs. Utilizing Haskell functions is presumed to have no side effects, as the evaluation order of the functions is independent \cite{c5}. Modern functional languages are characterized by being strongly typed, concise, clear, lazy, and easy to insure correctness. With no doubt, developing hardware circuits based on the functional programming paradigm is a promising and modern topic under investigation \cite{c2,c3,c4}. Much research effort has been done to benefit from the advantages of functional programming languages in hardware design including \textit{Lava} \cite{sheeran2005hardware}, \textit{Hawk} \cite{r1,r2}, \textit{Hydra} \cite{r4}, \textit{HML} \cite{r7}, \textit{MHDL} \cite{r6}, \textit{DDD} system \cite{r8}, \textit{SAFL} \cite{r10}, \textit{MuFP} \cite{r11}, \textit{Ruby} \cite{r12}, and \textit{Form} \cite{r14}.

HTCC compiles a subset of Haskell to Handel-C, in addition to automatically generating VHDL, Verilog, EDIF, and SystemC. The design of HTCC compiler includes lexical, syntax and semantic analyzers. The compiler is generated using ANTLR based-on a subset of Haskell grammar. HTCC Integrated Development Environment (IDE) produces a variety of analysis and schematic files. HTCC successfully connects to external tools, such as, DK Design Suite, Altera Quartus, and ModelSim. The developed compiler targets several FPGA types, and Altera DE2-70 and DE4 FPGA boards. The targeted area of application is cryptography, namely, the XTEA cipher.

The paper is organized so that Section~\ref{Introduction} presents the rapid prototyping methodology adopted by HTCC. Section~\ref{CCM} details the HTCC construction including the compiler and IDE designs. The compiler implementation is presented in Section \ref{Compiler_Implementation}. Sections~\ref{Haskell_Functions} and ~\ref{XTEA} present the compilation approach of first-class and higher-order functions and a case-study from cryptography. A thorough analysis and evaluation is presented in Section~\ref{Evaluation}. Section~\ref{conclusions} concludes the paper and sets the ground for future works.

\section{Background} \label{Introduction}
HTCC adopts the transformational derivation and refinement methodology of Abdallah et. al \cite{c10,c21}. The adopted methodology refines functional specifications into parallel hardware implementations in Handel-C. Several case-studies for the methodology were carried out by Damaj et. al \cite{c12,Damaj2011b,Damaj2009,Damaj2007a}, however the implementations did not include a compiler that automates the refinement procedure. 

\begin{figure}[!h]
	\centering
	\includegraphics[width=0.5\textwidth]{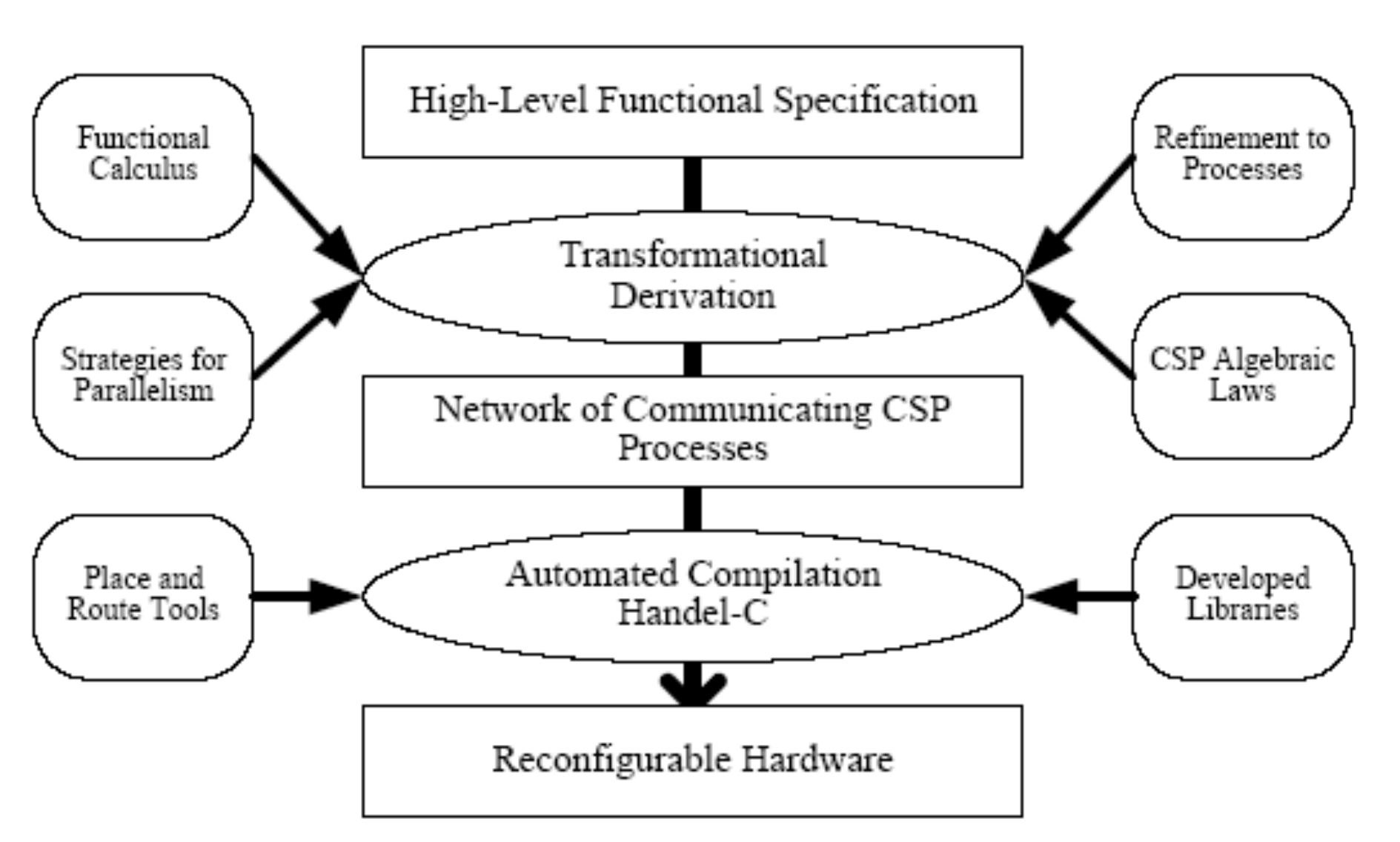}
	\caption{The transformational derivation and refinement methodology.}
	\label{RDM}
\end{figure}

Figure~\ref{RDM} depicts the step-wise refinement procedure, where functional specifications are refined to hardware. The adopted methodology is systematic in the sense that it is carried out using the following step-by-step procedure: 

\begin{itemize}
	\item Specify the algorithm in a functional setting relying on
	higher-order functions as the main building constructs wherever
	necessary.
	
	\item Apply the predefined set of rules to create the corresponding
	\textit{CSP} networks according to a chosen degree of parallelism.
	
	\item Write the equivalent \textit{Handel-C} code and complete the
	hardware compilation.
\end{itemize}

The refinement steps are aided by different compilers and integrated development environments. HTCC automates the development process including the background run of existing FPGA vendor interfaces and Haskell, Handel-C, VHDL, Verilog, EDIF, and SystemC compilers.

The adopted methodology refines both datatypes and functions. Datatypes are refined to \textit{Items}, \textit{Streams},
and \textit{Vectors} to create communicating entities based-on the message passing technique. The \textit{Item} corresponds to a basic type, such as an Integer data type , and it is to be communicated on a single communicating channel. The \textit{Stream} is a purely sequential method of communicating a list of values. The \textit{Vector} is a refinement of a simple list of items that communicates the entire structure in parallel \cite{Damaj2007a}.

In addition, the methodology refines functions to communicating processes. The refinement comprises a library of standard processes, such as, \textit{Produce} and \textit{Store} that aid the communication of refined datatypes. The \textit{Produce} process is used to produce values on the channels of a certain communication construct (\textit{Item}, \textit{Stream}, \textit{Vector}, etc.). These values are to be received and manipulated by another processes. The process \textit{Store} stores a communication construct in a simple or composite variable \cite{Damaj2007a}. 

The methodology also supports a rich set of refined higher-order functions, such as, \textit{map}, \textit{zip}, \textit{zipwith}, etc. The refinement of higher-order functions to processes could be done in stream or vector settings, or a combination of them. In Handel-C, datatypes are refined to structures (\textit{struct}), while processes are refined to \textit{macro procedures} \cite{Damaj2007a}. Handel-C compiler generates the required hardware circuits that can be mapped onto FPGAs.

\section{Compiler Construction} \label{CCM}
HTCC is a compiler that automates the presented refinement methodology. The presented version of HTCC Integrated Development Environment (IDE) supports the following:
\begin{itemize}
	\item Compiles a subset of Haskell to Handel-C
	\item Automatically connects to the DK Design Suite from Mentor Graphics to run the Handel-C Compiler; it verifies, generates, and analyzes the corresponding VHDL, Verilog, EDIF, or SystemC code
	\item Automatically connects to Glasgow Haskell Compiler (GHC) to run and test the Haskell code
	\item Automatically connects to Altera Quartus II to run, test, analyze hardware designs; place and route; produce bit files; and target specific FPGAs and FPGA boards.
	\item Provides an easy-to-use, rich, and modern development environment
\end{itemize} 

\subsection{Compiler Design using ANTLR}
HTCC is developed using the compiler-compiler tool ANTLR. ANTLR provides an easy-to-use compiler construction structure; ANTLR is efficient, reliable, and effective \cite{c12}. ANTLR uses an adaptive parsing technique that provides runtime grammar analysis \cite{c13}. Moreover, ANTLR uses the Extended Backus–Naur Form (EBNF). The efficiency and effectiveness of utilizing ANTLR is primarily due to its ability to support direct left-recursion, side-effecting actions (mutators) and predictions from the corresponding grammar \cite{c20}.

\begin{figure}[h!]
	\centering
	\includegraphics[height=2.5in, width=3.2in]{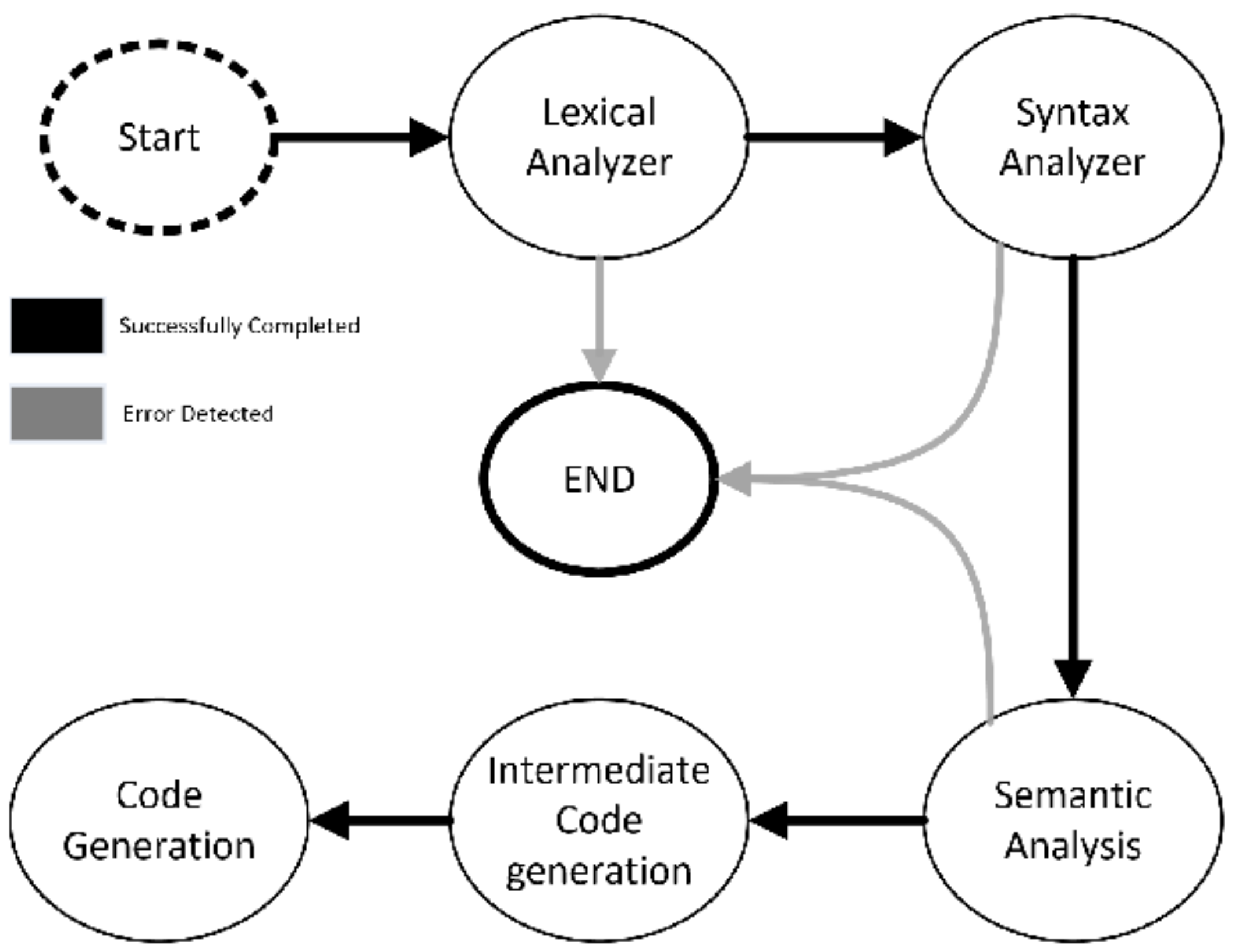}
	\caption{HTCC compiler state machine.}
	\label{SM}
\end{figure}

Figure~\ref{SM} demonstrates the state machine diagram of HTCC compilation procedure. The Lexical Analyzer analyzes the input Haskell code by producing a numbered list of lexemes. In addition, the Lexical Analyzer divides the code based on the provided grammar to prepare it for the syntax analysis. The Lexical Analyzer removes all white space between tokens and ignores any input with comment symbol "--".

The syntax analyzer is also generated using ANTLR, where a new parse tree is constructed every compilation. ANTLR provides the required Java library to construct parse trees and to walk through them starting on the leftmost side. During the walk-through, the program being compiled is checked for any errors based-on the provided grammar to ANTLR.

The third stage of HTCC compiler is the semantic analysis, where all types of all functions are checked and stored in a table for further processing. Semantic Analysis checks the types of inputs and outputs of each function. The semantic analyzer walks through the parse tree nodes using ANTLR's tree walker. If any datatype is found to be not supported or mismatched, HTCC terminates the compilation processes and reports the error. 

After a successful semantic analysis check, HTCC continues to the intermediate code generation and then to the final code generation. In the intermediate stage, all input and output interface buses and macros are generated. Then, the number of connections among macros is determined and passed to the final generation stage. During the final compilation stage, both Handel-C bus interfaces and Handel-C main method are generated. Moreover, the connections among all macros are generated. The current version of HTCC does not include an optimization stage. 

\begin{figure}[h!]
	\centering
	\includegraphics[height=3.2in, width=3.2in]{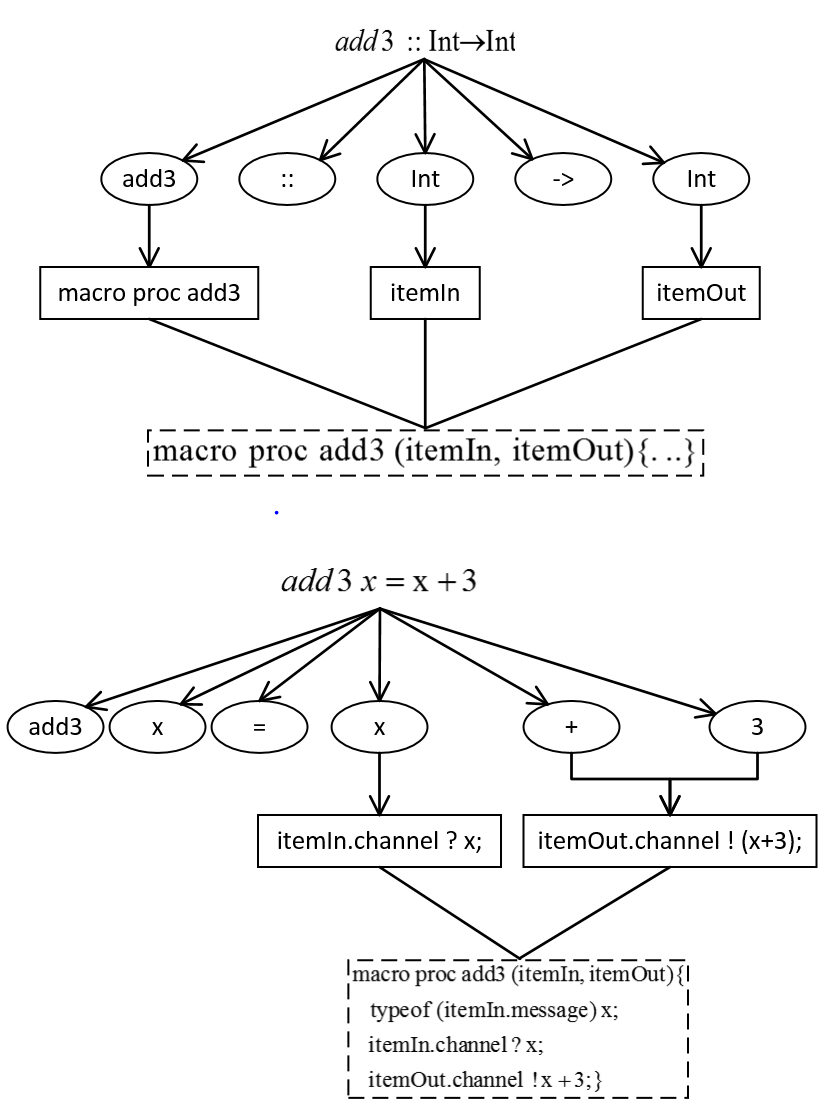}
	\caption{Code generation of items}
	\label{CG}
\end{figure}

Figure~\ref{CG} depicts the correspondence used to generate Handel-C macros from Haskell functions. An example Haskell function is as follows:
\begin{center}
	$add3::Int\rightarrow Int$\\
	$add3~x=~x+3$\\
\end{center}

The \textit{add3} function has one input and one output, where both are of type \textit{integer}. The corresponding Handel-C macro for \textit{add3} is as follows:

{\normalsize \begin{verbatim}
	macro proc add3 (itemIn, itemOut){
	typeof itemIn.message x;
	itemIn.channel ? x;
	itemOut.channel ! x+3;
	}
	\end{verbatim}}

\begin{figure}[h!]
	\centering
	\includegraphics[height=3.2in, width=3.2in]{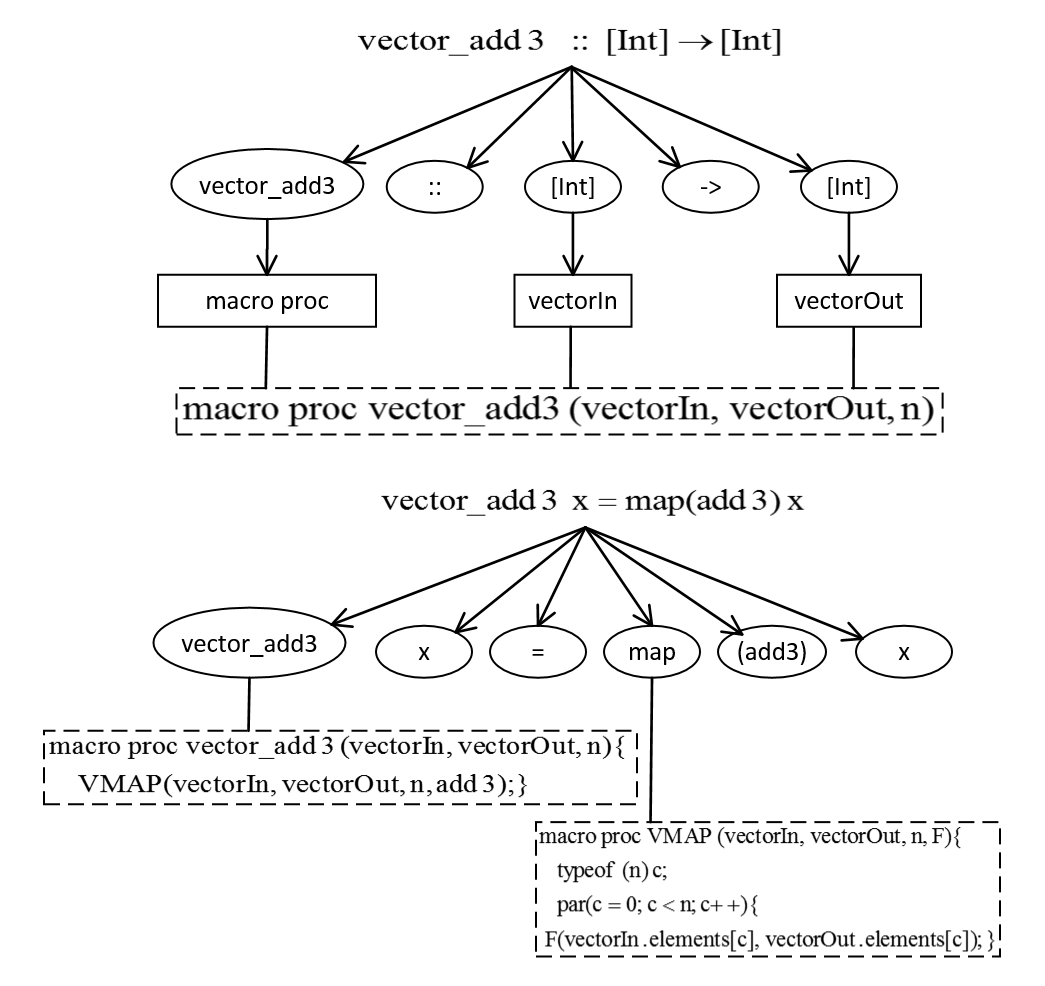}
	\caption{Code generation of parallel list processing}
	\label{CG2}
\end{figure}

It is very important to notice that \textit{add3} function can be utilized for list processing. The generation correspondence is shown in Figure~\ref{CG2}.
\begin{center}
	$vector\_add3::\lbrack Int\rbrack \rightarrow \lbrack Int\rbrack$\\
	$vector\_add3~x=~map(add3)~x$
\end{center}
The corresponding Handel-C code includes a version of \textit{add3} based on \textit{items}; the generic implementation of the parallel version of the higher-order function \textit{map} (VMAP); the implementation of function \textit{vector\_add3} that invokes \textit{VMAP} macro; and a main function that calls \textit{vector\_add3} with its inputs, outputs, and the number of elements in each vector. The parallel instances of add3 are replicated using the \textit{par} operator in Handel-C. The generated code is as follows:

{\normalsize \begin{verbatim}
	macro proc add3 (itemIn, itemOut){
	typeof itemIn.message x;f
	itemIn.channel ? x;
	itemOut.channel ! (x+3);}
	
	macro proc VMAP(vectorIn,vectorOut,n,F){
	typeof(n) c; 
	par(c=0;c<n;c++){
	F(vectorIn.elements[c], 
	vectorOut.elements[c]);}}  
	
	macro proc vector_add3 (vectorIn,vectorOut,n){ 
	VMAP(vectorIn,vectorOut,n,add3);
	}
	
	void main (){
	..
	vector_add3(vector0,vector1,5);
	..
	}
	\end{verbatim}}

\subsection{IDE Design}
\begin{figure}[h!]
	\centering
	\includegraphics[height=2.5in, width=3.1in]{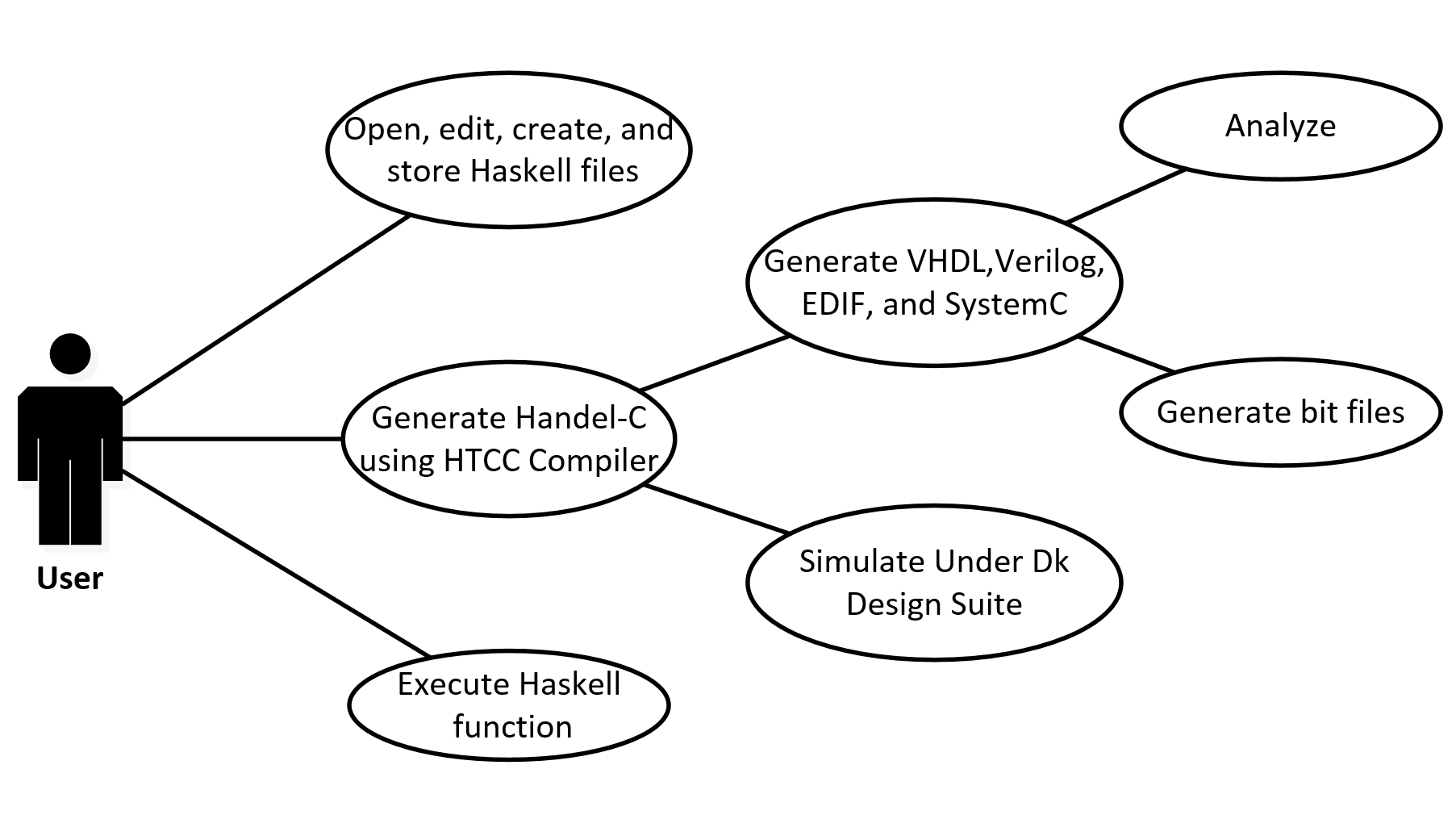}
	\caption{Use-Case diagram}
	\label{cased}
\end{figure}

The  technique used in the development of the IDE separates the programming concern in structuring the code in different Jar files. HTCC IDE adopts the iterative and incremental design model (IIDM) \cite{jacobson1999unified}. In the IIDM, each component of the IDE is developed separately as a standalone project which allows it to be integrated into multiple projects. The IDE is implemented using Java under Netbeans \cite{heffelfinger2015java}. The code editor is implemented using RSyntaxTextArea Java framework. The IDE theme is implemented using JTattoo Java framework. Figure~\ref{cased} demonstrates the use-case diagram of HTCC IDE. The proposed IDE supports the following:
\begin{itemize}
	\item Editing and storing project files
	\item Highlighting and automatic code completion
	\item File navigation, and allows to open multiple files simultaneously
	\item Running Haskell code under GHC
	\item Compiling Haskell code to Handel-C code. Accordingly simulating Handel-C code and generating VHDL, EDIF, Verilog, and SystemC implementations.
	\item Compiling the generated HDL files using Altera Quartus. Accordingly, producing analysis and FPGA mapping files.
\end{itemize}

\begin{figure}[h!]
	\centering
	\includegraphics[height=3in, width=3.5in]{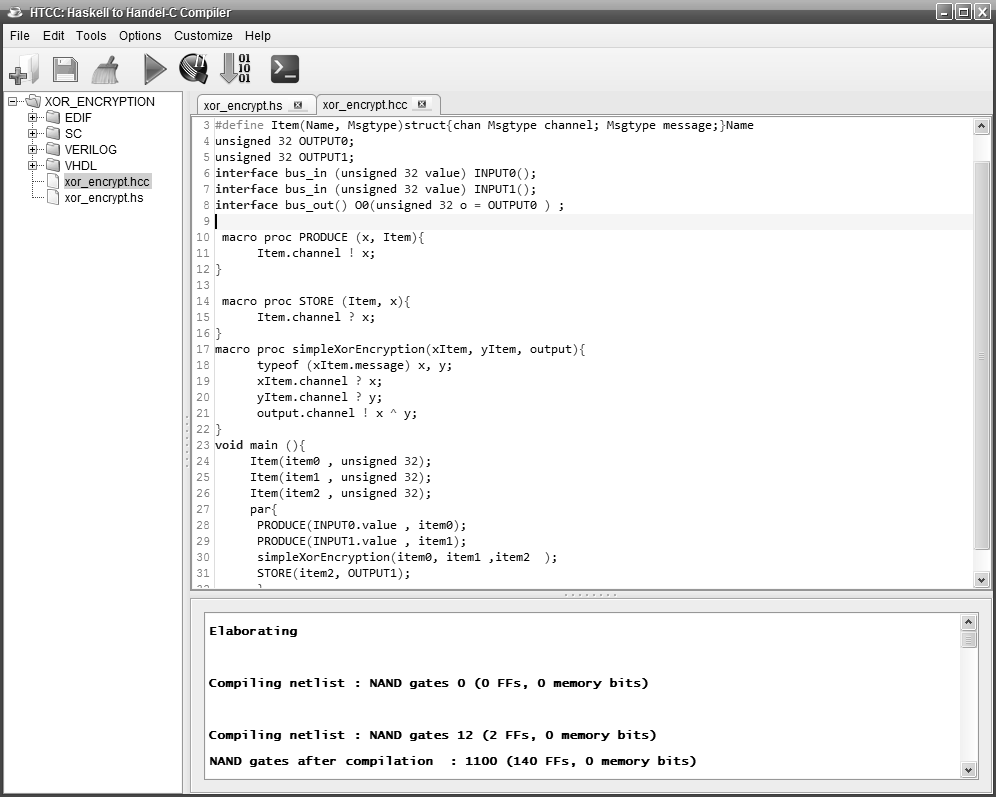}
	\caption{HTCC IDE}
	\label{IDE}
\end{figure}

The IDE connects HTCC Compiler to external tools, such as, DK Design Suite to simulate and generate VHDL, Verilog, EDIF, and SystemC files. In addition, the IDE connects the compiler to Altera Quartus using the TCL commands to synthesize and generate timing analyses, pin assignments for FPGA boards, and generate bit files to program the targeted FPGAs. GHC is also connected to the IDE to execute and verify Haskell functions. Figure~\ref{IDE} shows a snapshot of the HTCC IDE.

\section{Compiler Implementation} \label{Compiler_Implementation}

The following subset of Haskell grammar is part of HTCC compiler code. Here, functions are divided into decelerations (\textit{dcFun}) and definitions (\textit{dFun}):
\\
\\
\\
\\
\\
\\
\\
\\
\\
\\
\\
\\
PROG : $STAT+;$\\
STAT : $dcFun;$\\
dcFun : $ID~'::'~formalType (->) * NL+ dFun;$\\
expr : $expr~op=('*'| '/' )(DIGIT|expr )$\\
$~~~~~~~~|expr op=('.\&.' | '.||.')(DIGIT|expr )$\\
$~~~~~~~~|expr op=('+'| '-')(DIGIT|expr )$\\
$~~~~~~~~|('xor' expr DIGIT)$\\
$~~~~~~~~|('shiftL' expr DIGIT)$\\
$~~~~~~~~|('shiftR' expr DIGIT)$\\
$~~~~~~~~|mPassing (mPassing)*$\\
$~~~~~~~~|expr mPassing$\\
$~~~~~~~~|ID*$\\

According to the proposed grammar an expression (\textit{expr}) has multiple meanings that captures the definition of the function. \textit{expr} can be any arithmetic or logic operation between two or more variables. In addition, an expression \textit{expr} can call other functions that take place at \textit{mPassing} node. Figure 7 demonstrates the parse tree of the following function:
\begin{center}
	$f~::~Int\rightarrow Int$\\
	$f~x=x~+~3$
\end{center}
\begin{figure}[h!]
	\centering
	\label{parsetree}
	\includegraphics[height=2.3in, width=3.2in]{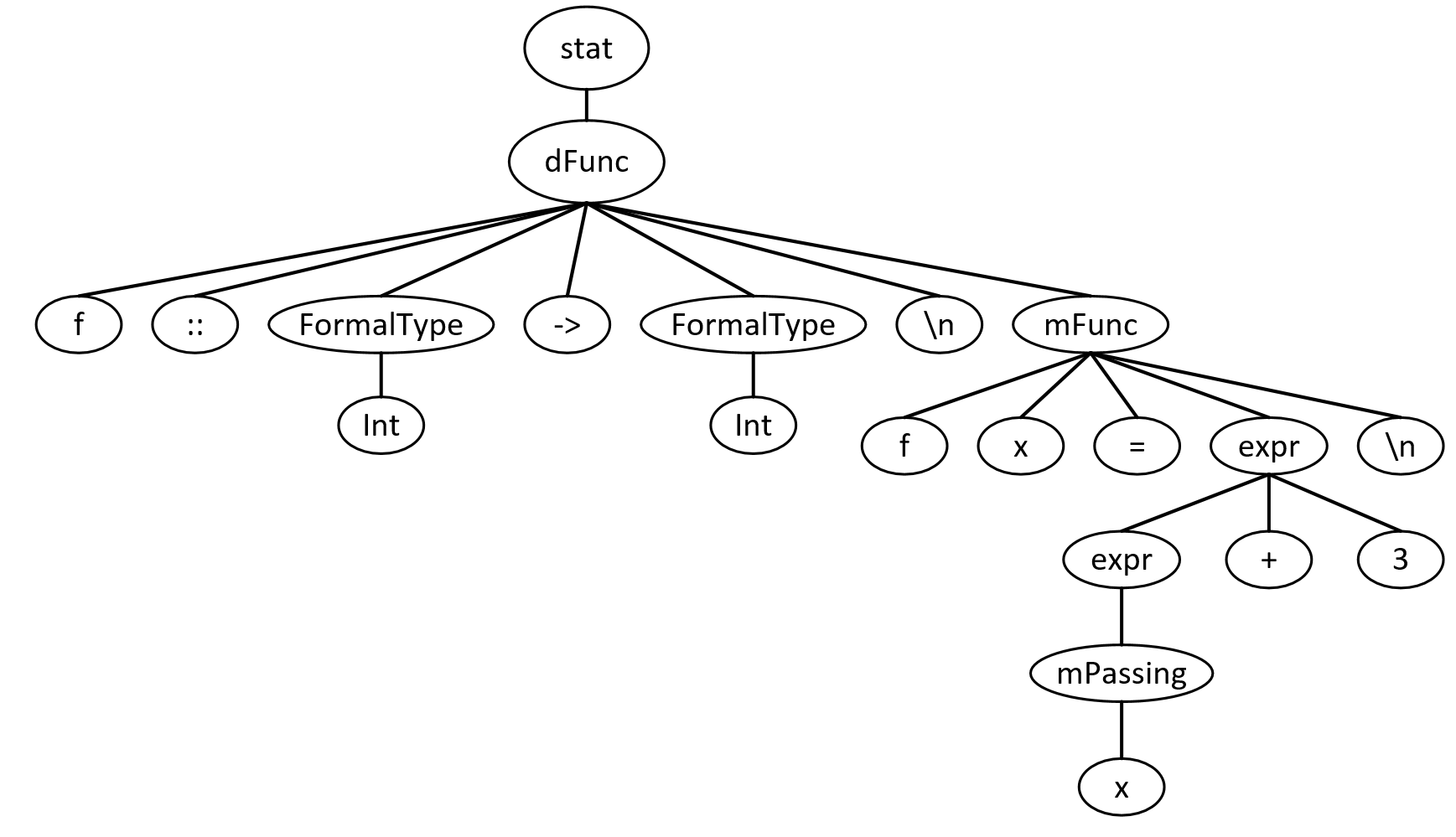}
	\caption{The parse tree of function f.}
\end{figure}
A subset of the lexer grammar is as following:\\
ID	:$~[a-zA-Z]^+~ [0-9]^*;$\\
NL	:$~'\backslash r'~?~ '\backslash n' ; $\\
ARROW   :$~'- >'~|~ '\rightarrow';$\\
WS	:$~[ \backslash t]^+ \rightarrow SKIP;$\\
DIGIT	:$~[0-9]^+; $\\
COMMENT :$~'--'~~ .*?~ ~'\backslash r'~? ~'\backslash n' \rightarrow SKIP ;$\\

\section{FIRST-CLASS AND HIGHER-ORDER \\ HASKELL FUNCTIONS} \label{Haskell_Functions}
HTCC can generate both first-class and higher-order functions. First-class functions represent simple binary operations, while higher-order functions can take other functions as parameters and usually are operated on lists. 

\subsection{First-Class Functions}
A sample generation of the binary operation OR is shown in the following: 
\begin{center}
	$or::Int\rightarrow Int\rightarrow Int$\\
	$or~~a~~b = a~ .|. ~b$
\end{center}
By compiling the function \textit{or} under HTCC, the generated Handel-C code comprises three items - each has a message of width 32 bits. The first two items are \textit{a} and \textit{b}, and the third item is where the result is stored. In addition, HTCC generates the macro \textit{OR}. HTCC generates three interfaces that are \textit{input0}, \textit{input1}, and \textit{output0} for the inputs and output. In the main method, HTCC creates three items to produce the two inputs and store the output. Similar first-class functions, such as, \textit{AND}, \textit{XOR}, \textit{ADD}, \textit{SUB}, \textit{DIV} can be generated in a similar way. To run the compiled code on the Altera DE2-70, the following is automatically generated by HTCC.

{\footnotesize \begin{verbatim}
set clock = external"AD15";
set reset = external"L8";
#define Item(Name, Msgtype)struct{chan Msgtype 
channel; Msgtype message;}Name

unsigned 32 OUTPUT0;
interface bus_in (unsigned 32 value) INPUT0();
interface bus_in (unsigned 32 value) INPUT1();
interface bus_out() O0(unsigned 32 o = OUTPUT0 ) ;

macro proc OR (xItem, yItem, 
itemOut){
typeof (xItem.message) x,y;
item0In.channel ? x;
item1In.channel ? y;
itemOut.channel ! x || y;}

void main (){
Item(item0 , unsigned 32);
Item(item1 , unsigned 32);
Item(item2 , unsigned 32);
par{
PRODUCE(INPUT0.value , item0);
PRODUCE(INPUT1.value , item1);
OR(item0, item1, item2  );
STORE(item2, OUTPUT0);}}

\end{verbatim}}

\subsection{Higher-Order Functions}
HTCC utilizes a set of parallel and sequential versions of a set of higher-order functions including \textit{map}, \textit{zipWith}, \textit{foldr}, etc. The following is a sample generation of a parallel zipping of two lists with multiplication. Each list contains ten elements. The generation employs the VectorOfItems structure and the parallel version of produce and store macros.\\

\noindent $mul~::~Int\rightarrow Int\rightarrow Int$\\
$mul~x~y=x*y$\\
\\
$two\_vectors\_mul~::\lbrack Int\rbrack\rightarrow \lbrack Int\rbrack\rightarrow \lbrack Int\rbrack$\\
$two\_vectors\_mul~a~b=zipWith(mul)~a~b$
{\small \begin{verbatim}
	macro proc mul (xItem, yItem,output){
	typeof (xItem.message) x, y;
	xItem.channel ? x;
	yItem.channel ? y;
	output.channel ! (x*y);}
	
	macro proc VZIPWITH ( vectorIn1, vectorIn2, 
	vectorOut, n, F){
	typeof (n) c;
	par (c =0; c< n; c++){
	F(vectorIn1.elements[c], vectorIn2.elements[c],
	vectorOut.elements[c]); }}
	
	macro proc two_vectors_mul(vectorIn1,vectorIn2,
	vectorOut,n){
	VZIPWITH(vectorIn1, vectorIn2, vectorOut, 100, mul);}
	
	void main (){
	VectorOfItems(vector0, 10, unsigned 32);
	VectorOfItems(vector1, 10, unsigned 32);
	VectorOfItems(vector2, 10, unsigned 32);
	par{
	VPRODUCE(INPUT0, vector0, 10);
	VPRODUCE(INPUT1, vector1, 10);
	two_vectors_mul(vector0,vector1,vector2,10);
	VSTORE(vector2, OUTPUT0);}}
	\end{verbatim}}

\section{Case-Study: The Rapid Prototyping of XTEA under HTCC} \label{XTEA}
To test the applicability of the developed compiler, we use the extended tiny encryption algorithm (XTEA) as a case-study. XTEA uses a 128-bit key to encrypt a 64-bit block ciphertext which follows Feistel cipher’s structure with a variable number of rounds. The 128-bit plaintext is divided into two integers \textit{V0} and \textit{V1}. The key produces a set of integer sub-keys to be distributed to the appropriate round. XTEA is small in size, light in weight, low in power, and a secure block cipher \cite{c14}. The following is the functional specification of the XTEA single round under Haskell:\\
\\
$xteasround::Int\rightarrow uInt32\rightarrow(uInt32,uInt32)\rightarrow uInt32\rightarrow(uInt32,uInt32)$\\
$xteasround~1~sum~x@(v0,v1)~key0=x$\\
$xteasround~rounds~sum~(v0,v1)~key0=xteasround~$\\
$~~~(rounds+1) ~new\_sum~(new\_v0,new\_v1)~key~where$\\
$~~~~~~~new\_v0 = xteav0~v0~v1~sum~key0$\\
$~~~~~~~new\_sum= xteasum~sum$\\
$~~~~~~~new\_v1 = xteav1~new\_v0~v1~new\_sum~key0$\\
\\
$xteav0~~::~uInt32\rightarrow uInt32\rightarrow uInt32 \rightarrow uInt32 \rightarrow uInt32$\\
$xteav0~v0~v1~~sum~key0=~~v0~+$\\
$(xor~(key0+sum)~(v1+(xor~(shiftL~v1~4)~$
$(shiftR~v1~5)))$\\
\\
$xteasum~::~uInt32 \rightarrow uInt32$\\
$xteasum~sum=sum+0x9e3779b9$\\
\\
$xteav1~~::~uInt32\rightarrow uInt32\rightarrow uInt32 \rightarrow uInt32 \rightarrow uInt32$\\
$xteav1~v0~v1~sum~key0=~~v1~+$\\
$(xor~(key0+sum)~(v0+(xor~(shiftL~v0~4)~$
$~(shiftR~v0~5)))$\\

The data type \textit{uInt32} is a user-defined unsigned integer with 32 bits width. A single round of XTEA generates the following sample main function. However, the function \textit{xteasround} produces a macro \textit{XTEASROUND} when the 32 rounds are replicated to implement the top-level function \textit{xtea}. 

\begin{verbatim}
void main {
par{
PRODUCE(INPUT0.value, item0);
PRODUCE(INPUT1.value, item1);
PRODUCE(INPUT2.value, item2);
PRODUCE(INPUT3.value, item3);
xteav0(item0, item1, item2, item3, item4);
xteasum(item3, item5);
xteav1(item4, item1, item2, item5, item6);
STORE (item4, OUTPUT0);
STORE (item5, OUTPUT1);
STORE (item6, OUTPUT2);}}
\end{verbatim}

\begin{figure}[h!]
	\centering
	\includegraphics[height=3.1in, width=2in]{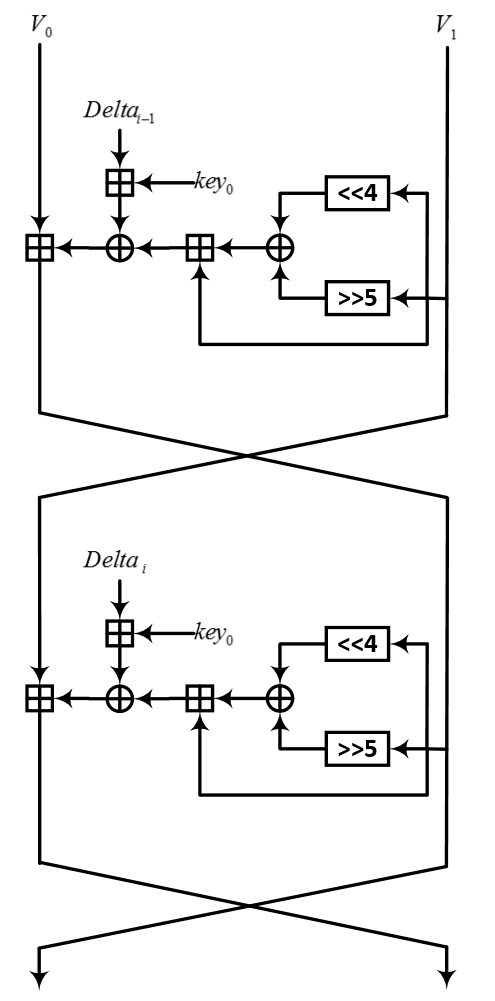}
	\caption{A single XTEA round with its internal computational constructs. The crossed square for the sum, crossed circle for an XOR, $>>$ for a right shift, $<<$ for a left shift.}
	\label{XTEASINGLE}
\end{figure}

\section{Analysis and Evaluation} \label{Evaluation}
The proposed compiler allows for the rapid prototyping of hardware circuits at a high-level of abstraction based-on functional specifications. Functional programming enables designing hardware using clear, concise, and correct-by-construction specifications. Overall, the proposed compiler translates a subset of Haskell to Handel-C and thus enables the usage of Haskell as a hardware description language for programming FPGAs.

HTCC adopts an effective transformational derivation approach that enables the systematic development of CSP concurrency descriptions. Accordingly, the automatic generation of Handel-C code is possible and effective in generating VHDL, EDIF, Verilof, and SystemC descriptions. The refinement methodology provides a variety of parallelism techniques to specify the required degree of parallelism. The methodology provided HTCC with the characteristics of generating a variety of implementations with different parallel characteristics. HTCC benefited from the off-the-shelf first-order, higher-order, and application-specific libraries provided by Damaj et al. \cite{Damaj2007a,Damaj2009,Damaj2011b} and automated the refinement procedure.     

HTCC IDE enables the testing and evaluation of both Haskell and Handel-C code through the background connection to their native compilers. HTCC IDE offers the options to display analysis reports supported by Quartus, such as, power consumption, area utilization, timing, RTL views, pin assignments, etc. Furthermore, the adopted IIDM technique allows for the rapid development and integration of the various parts of the IDE with simplicity. 

Although the use of ANTLR made the compiler implementation simple, additions are necessary. The main addition in HTCC is the semantic analyzer that was embedded into the adopted ANTLR structure. The embedding enabled effectively for type checking and error reporting using the supported exception handling mechanism.

Table~\ref{table1} presents the performance analysis results of the XTEA cipher as generated by HTCC and tested under Cyclone II, Stratix IV and Virtex-6 FPGAs. The Cyclone II FPGA is part of the targeted DE2-70 board. The Stratix IV FPGA is part of the targeted Altera DE4 board. The Virtex-6 FPGA is a high-speed FPGA from Xilinx. The Total Number of NAND Gates as measured under DK Design Suite is 467969 with a total of 192 clock cycles. The highest frequency achieved is 648.54 MHz under Virtex-6, and the lowest power consumption achieved is 219.62 mW under the Cyclone II. In addition, the highest throughput is 219.3 Mbps under Xilinx Virtex-6 FPGA.

\begin{table}[h!]
	\centering
	\caption{XTEA Implementation Results}
	
	\label{table1}
	\resizebox{8.8cm}{!}{\begin{tabular}{l || l || l || l}
		\hline
		\textbf{}        & \textbf{Cyclone II} & \textbf{Stratix IV} & \textbf{Virtex-6} \\ \hline
		\textbf{Total logic elements}   &   15,573 LE  & 1221 ALUTs &26660  Slices     \\ \hline
		\textbf{Fmax} (MHz)                 & 183.18    & 513.8  & 648.54      \\ \hline
		\textbf{Total Execution Time} (ns)  & 5.46    & 1.95 & 1.52      \\ \hline
		\textbf{Throughput} (Mbps)           & 61.06     & 171.26  & 219.3   \\ \hline
		\textbf{Power consumed} (mW)            & 219.62     & 888.47 & 912.4   \\ \hline
	\end{tabular}}
\end{table}

As compared to the performance reported in \cite{c14,c15,c16,c17}, the results produced by HTCC achieved the highest throughput of 219.3 Mbps under the Virtex-6 (See Table~\ref{t3}). A behavioral implementation of the XTEA cipher under VHDL achieved 134 Mbps, however, the main purpose of the implementation was to achieve a compact and low-power design \cite{c14}. The manual Handel-C (HC) implementation achieved a speed of 44.25 Mbps with an Fmax of 177 and an area of 720 Logic Elements.

\section{Conclusion} \label{conclusions}
HTCC is a Haskell to Handel-C hardware compiler that targets FPGAs. HTCC automates a transformational derivation methodology to rapidly produce hardware circuits from functional specifications. The adopted methodology refines functional programs to a formal concurrency framework, namely, CSP. The methodology enables the systematic refinement of the CSP descriptions to Handel-C; HTCC comes to make this process automatic. Nevertheless HTCC doesn't produce CSP descriptions, this is identified as a future development. The developed compiler effectively produces hardware circuits in various descriptions and languages, such as, VHDL, Verilog, EDIF, and SystemC. HTCC connects to a bouquet of hardware design tools to produce a rich-set of analysis reports and bit-stream files that can map to different FPGAs. The paper includes a case-study from cryptography that produces comparable, and in some instances better results than what is reported in the literature. Indeed, HTCC adopted a functional programming style to benefit from its simplicity, conciseness, and correctness. Future work includes expanding the area of application and widening the pool of implemented Haskell syntax and parallelization options.

\bibliographystyle{IEEEtran}
\bibliography{DSD}  
\newpage
\begin{table}[H]
	\centering
	\caption{Compassion among similar XTEA hardware implementation}
	\label{t3}
	\begin{tabular}{l || l || l || l || l}
		\centering
		\textbf{\textit{Reference}}        & \textbf{\cite{c15}} & \textbf{\cite{c16}} & \textbf{\cite{c17}} &\textbf{\cite{c14}}\\ \hline
		
		\textbf{Logic elements}   &      NA& 424 LUTs& 1182 LUTs    & 539 Slices\\ \hline
		\textbf{Fmax} (MHz)                 &    NA & NA& 71.11  & 142.4    \\ \hline
		\textbf{Total Exe. Time}   &    2,48 ms& NA& 14.06 ns  &  NA  \\ \hline
		\textbf{Throughput}            & 0.39 kB/s     & NA& NA & 134 Mbps \\ \hline
		\hline
			\textbf{\textit{Reference}}         & \textbf{Manual HC} & \textbf{HTCC} & & \\ \hline
			
			\textbf{Logic elements}   &      720 LE & 26660 Slices & &\\ \hline
			\textbf{Fmax} (MHz)                 &   177  & 648.54 & &  \\ \hline
			\textbf{Total Exe. Time}   &    5.6 ns & 1.52 ns & & \\ \hline
			\textbf{Throughput}            & 44.25 Mbps  & 219.3 Mbps & & \\ \hline

	\end{tabular}
\end{table}
\end{document}